\documentclass{emulateapj}
\shorttitle{Disk-warping planet b}
\shortauthors{Dawson et al.}
\usepackage{natbib}
\newcommand{\icarus}{Icarus\ }
\begin{document}


\title{On the misalignment of the directly imaged planet $\beta$ Pictoris b with the system's warped inner disk}
\slugcomment{In press ApJL; accepted: November 1, 2011}

\author{Rebekah I. Dawson}
          \email{rdawson@cfa.harvard.edu}
\author{Ruth A. Murray-Clay}
\affil{Harvard-Smithsonian Center for Astrophysics \\ 
       60 Garden Street, Cambridge, MA 02138 USA}

\author{Daniel C. Fabrycky}
\affil{UCO/Lick Observatory, University of California, Santa
  Cruz, CA 95064}
  
\begin{abstract}
The vertical warp in the debris disk $\beta$ Pictoris -- an inclined inner disk extending into a flat outer disk -- has long been interpreted as the signpost of a planet on an inclined orbit. Direct images spanning 2004-2010 have revealed $\beta$ Pictoris b, a planet with a mass and orbital distance consistent with this picture. However, it was recently reported that the orbit of planet b is aligned with the flat outer disk, not the inclined inner disk, and thus lacks the inclination to warp the disk. We explore three scenarios for reconciling the apparent misalignment of the directly imaged planet $\beta$ Pictoris b with the warped inner disk of $\beta$ Pictoris: observational uncertainty, an additional planet, and damping of planet b's inclination. We find that, at the extremes of the uncertainties, the orbit of $\beta$ Pictoris b has the inclination necessary to produce the observed warp. We also find that if planet b were aligned with the flat outer disk, it would prevent another planet from creating a warp with the observed properties; therefore planet b itself must be responsible for the warp. Finally, planet b's inclination could have been damped by dynamical friction and still produce the observed disk morphology, but the feasibility of damping depends on disk properties and the presence of other planets. More precise observations of the orbit of planet b and the position angle of the outer disk will allow us to distinguish between the first and third scenario.
\end{abstract}


\section{Introduction}
\label{sec:intro}
The $\beta$ Pictoris debris disk is a rich system, with observational features resulting from the interplay of gravity, radiation pressure, collisions, infalling comets, sculpting by planets, and the physical properties of the gas, dust, and rocks that comprise the disk. In the quarter-century following its discovery \citep{1984S}, $\beta$ Pictoris has epitomized young planetary systems, amenable to state-of-the-art observations and to modeling of planetary formation processes. For a review, see \citet{1998V}. 

A striking vertical warp in the $\beta$ Pictoris disk, at approximately 85 AU from the star, appears in optical and near-infrared images \citep[e.g.][]{2000H,2006G}. \citet{1997M} and \citet{2001A} demonstrated that a perturbing planet on an inclined orbit could produce the warp, whose distance constrains the posited planet's mass and position. A decade later, \citet{2009L,2010L} discovered, via direct imaging, $\beta$ Pictoris b, a planet consistent with producing the warp, if correctly inclined relative to the disk.

However, \citet{2011C} -- stitching together recent \citep{2010Q,2011BL} data and newly reduced data \citep[collected by][]{2010L}  -- recently measured the planet's astrometric orbit and reported it to be, surprisingly, misaligned with the warp. They urged revisiting whether planet b could produce the warp and suggested an undiscovered additional planet as an alternative culprit.

Here we explore three scenarios to reconcile the apparent misalignment of planet b with $\beta$ Pictoris's warped inner disk. In Section \ref{sec:model}, we model an inclined planet sculpting a planetesimal disk. Then (Section \ref{sec:sculpt}), we consider the first scenario: within the extremes of the observational uncertainties, $\beta$ Pictoris b has the inclination to produce the warp. In Section \ref{sec:planetc}, we evaluate the possibility that another planet created the warp; however, we find that the presence of planet b on a flat orbit prevents another object from creating the observed warp. In Section \ref{sec:damp}, we explore the third possibility: planet b had a higher inclination in the past, created the warp, and then its inclination damped. Thus we suggest (Section \ref{sec:discuss}) that planet b produced the warp, whether or not its orbit is currently aligned with the warped inner disk. 

\section{Model of a debris disk sculpted by an inclined planet}
\label{sec:model}

\begin{figure*}[htbp]
\begin{centering}
\includegraphics{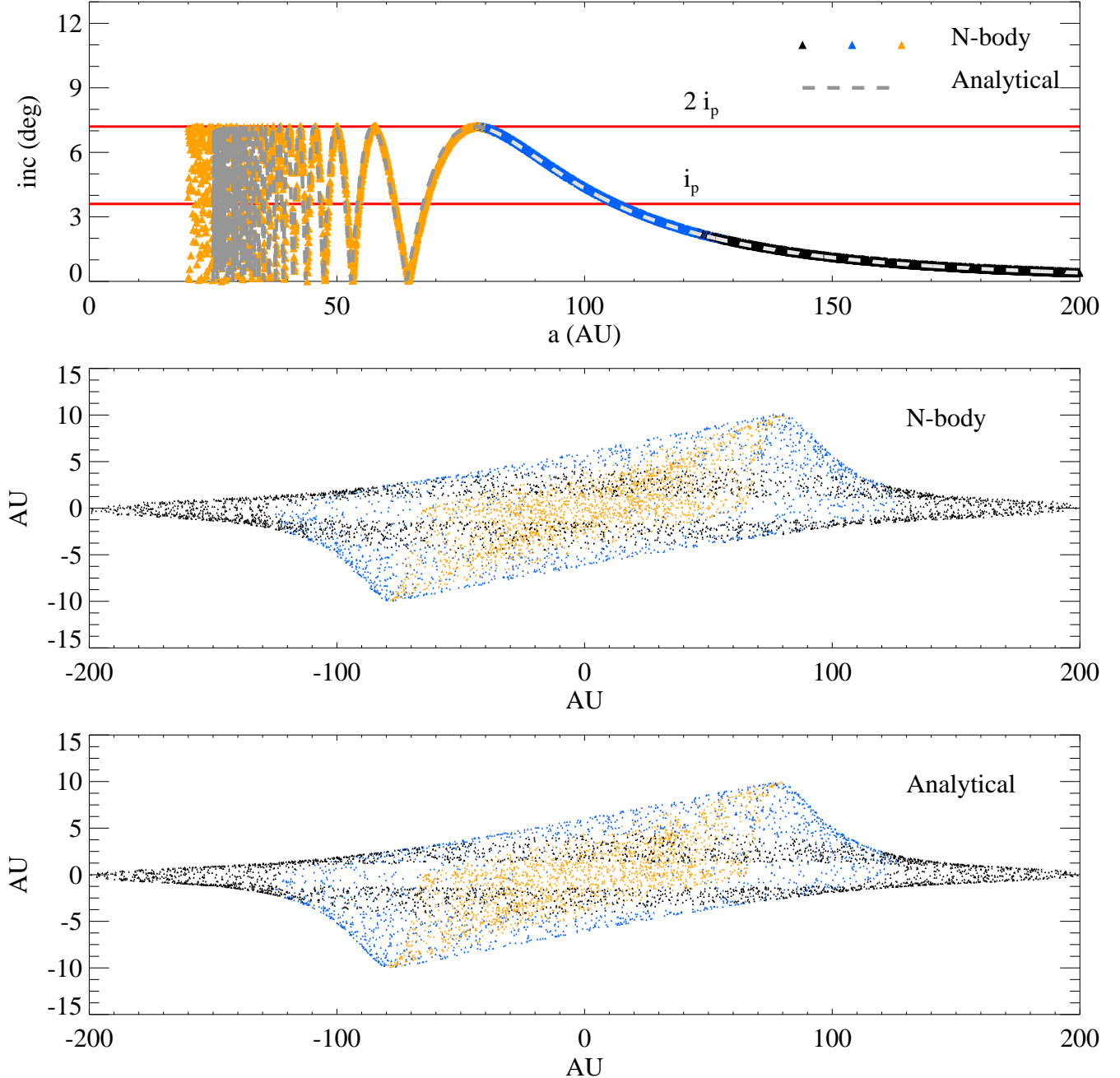}
\caption{Snapshot (8 Myr) from an N-body simulation and an analytical model of an inclined planet sculpting an initially flat planetesimal disk. Top: Planetesimal inclinations vs. semi-major axis. The red lines indicate the planet's inclination $i_p$ and $2i_p$. The orange, blue, and black line is composed of triangles, each marking the inclination of a corresponding planetesimal in the middle panel.  The planetesimals are color-coded: orange triangles have completed an oscillation, blue are just reaching their maximum inclination, and black are still at low inclinations. We use our analytical model (eqn. \ref{eqn:seci}) to calculate the dashed gray line. Middle: Projected positions of the planetesimals. Lower: Same as middle using the analytical model; it reproduces the N-body simulation well. Compare these two lower panels to images of the disk, such as those shown in \citet{2000H}, Fig. 8 and \citet{2006G}, Fig. 5.  \label{fig:warp}}
\end{centering}
\end{figure*} 

The warping of the $\beta$ Pictoris debris disk by a planet on an inclined orbit results from the secular evolution of the disk's component planetesimals. The planet secularly forces the disk, causing the planetesimals to oscillate about the planet's inclined plane. Planetesimals in an initially flat disk (inclination $i=0$ relative to the plane of the flat outer disk) reach a maximum $i$ of twice the planet's inclination $i_p$. The oscillation frequency decreases with the planetesimal's semi-major axis: in a young system like $\beta$ Pictoris, the inner planetesimal disk -- which secularly evolves quickly --- is centered on the planet's inclined plane, while the outer disk -- which secularly evolves slowly -- is still near its initial low inclination. Thus secular evolution produces an inclined inner disk aligned with the planet's inclined plane, a flat outer disk, and a warped feature, extending to $2i_p$, at the distance, $a_{\rm warp}$, where planetesimals are just reaching their maximum inclination. This distance constrains the disk's evolution time and the perturbing planet's mass and location. In Fig. \ref{fig:warp}, we plot the final inclinations and projected positions from an N-body simulation of an initially flat planetesimal disk sculpted by a planet on an inclined orbit.  Each of 6000 test-particle planetesimals begins with $e = i = 0$, a semi-major axis $a$ between 20-200 AU, and a random periapse, longitude of ascending node, and mean anomaly. The planet has the observed mass ($m_p = 9 m_{\rm Jupiter}$) and semi-major axis ($a_p = 9.5$ AU) of $\beta$ Pictoris b, and $i_p = 3.6^\circ$; the star has mass $m_* = 1.75 m_{\rm sun}$. We used the \emph{Mercury} 6.2 \citep{1999C} hybrid integrator with a step size of 200 days over a timespan of 8 Myr. The system's age, 12$^{+8}_{-4}$ Myr \citep{2001Z}, minus the planet formation time, $<3-5$ Myr \citep{2007H,2009CL}, imply that warp production likely began 3-20 Myr ago. We choose the viewing orientation to make both the outer disk and the planet's orbit edge-on, as observed.

\subsection{Planetesimal inclination evolution}

Secular evolution times set the warp's position: at the pointy part, planetesimals, in the midst of their first cycle about the planet's inclined plane, are just reaching their maximum $2i_p$ (blue triangles, Fig. \ref{fig:warp}). The components $p$ and $q$ of the planetesimal's instantaneous inclination, $i = \sqrt{p^2+q^2}$, evolve as:
\begin{eqnarray}
\label{eqn:seci}
p = i_{\rm free} \sin(f t + \gamma) + p_{\rm forced} \nonumber \\
q = i_{\rm free} \cos(f t + \gamma ) +q_{\rm forced}  \nonumber \\
f = -  \frac{n}{4} \frac{m_p}{m_*}  \alpha \bar{\alpha} b_{3/2}^{(1)}(\alpha)     \nonumber \\
\end{eqnarray}
where
\begin{eqnarray}
\alpha =  
   \left\{\begin{array}{l}
    a/a_p, \qquad\qquad\qquad\qquad\mbox{$a < a_p$;} \\  
    \rule{0ex}{5ex} a_p/a,\qquad\qquad\qquad\qquad\mbox{$a > a_p;$}   
  \end{array}\right.   \nonumber \\ \nonumber \\
  \bar{\alpha} =  
   \left\{\begin{array}{l}
    a/a_p, \qquad\qquad\qquad\qquad\mbox{$a < a_p$;} \\  
    \rule{0ex}{5ex} 1,\qquad\qquad\qquad\qquad\qquad\mbox{$a > a_p.$}   
  \end{array}\right.   \nonumber \\
\end{eqnarray}
The function $b$ is a standard Laplace coefficient \citep{2000M}, and $n = (Gm_*/a^3)^{1/2}$ is the planetesimal's mean motion. When a single inclined planet forces an initially cold disk, the forced plane, with inclination $i_{\rm forced} = \sqrt{p_{\rm forced}^2+q_{\rm forced}^2}$, is the inclined plane of the perturbing planet, $i_p$, relative to the flat outer disk. Thus a planetesimal's inclination, initially at $i = 0$, oscillates as:
\begin{eqnarray}
  \label{eqn:ivolve}
i = 2 i_p | \sin(ft/2)|  \nonumber \\
\end{eqnarray}

\subsection{Constraining the sculpting planet's orbit}
\label{subsec:constrain}
The disk reaches its maximal vertical extent, $z_{\rm warp}$, at $a_{\rm warp}$. From eqn. \ref{eqn:ivolve}, the warp's secular oscillation frequency $f$ is related to the disk's evolution time $\tau$ by $ f\tau = \pi$, constraining the mass and semi-major axis of the perturbing planet:
\begin{eqnarray}
\pi/\tau  =\frac{n_{\rm warp}}{4} \frac{m_p}{m_*} \alpha_{\rm warp} \bar{\alpha}_{\rm warp} b_{3/2}^{(1)}(\alpha_{\rm warp} ) 
\label{eqn:constrain}
\end{eqnarray}
\noindent and the perturbing planet's inclination is separately constrained by:
\begin{eqnarray}
\label{eqn:ip}
\tan(2 i_p) = \frac{z_{\rm warp}}{a_{\rm warp}} 
\end{eqnarray}

Our eqn. \ref{eqn:constrain} is equivalent to the warp condition in \citet{1997M}, Section 5 and \citet{2001A}, Section 1.2, derived from tidal theory. Additionally, \citet{1997M} modeled the disk's evolution using hydrodynamic simulations. However, using the secular equations above, we can not only constrain the parameters of a perturbing planet but produce the time-evolving disk morphology (e.g. Fig. \ref{fig:warp}, panel 3), facilitating comparisons to observations without simulations.

\begin{figure}[htbp]
\begin{centering}
\includegraphics{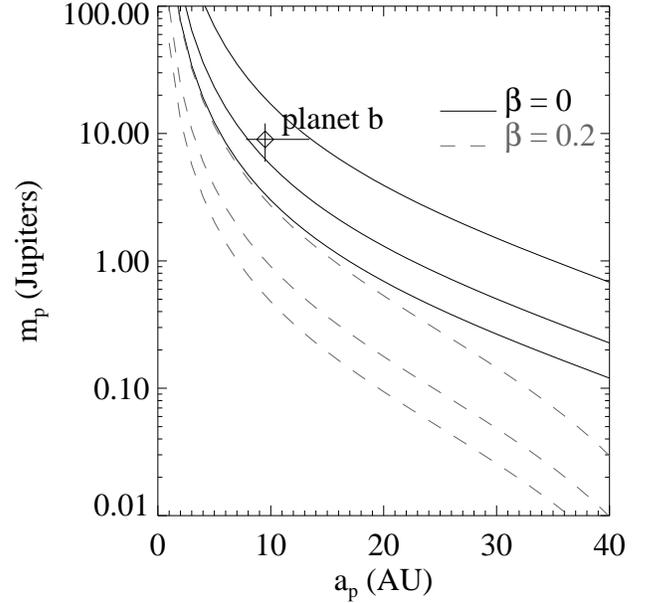}
\caption{Three curves (solid) of $(a_p,m_p)$ that produce a warp at 85 AU for (top to bottom) $\tau = $ 3, 9, 17 Myr. A planet with properties above the lowest curves significantly impacts the warp evolution over the system lifetime. Equivalent curves for observed dust grains with radiative forcing parameter $\beta = 0.2$ are plotted in dashed grey. Increasing $\beta$ decreases the warp distance in the planetesimal parent population, effectively shortening the warp propagation time. \label{fig:planetb}}
\end{centering}
\end{figure}

The warp-revealing visual observations measure the light scattered by \emph{dust}, likely produced by recent collisions of the planetesimal parent bodies. These dust grains are subject to radiative pressure, which induces larger, eccentric orbits relative to their parent bodies. Thus radiation pressure effectively increases the distance of the warp by a factor of $\sim 1/(1-2\beta)$, where $\beta$ is the ratio of the radiation forces to gravitational forces \citep[see][]{2009C}. For example, if $\beta = 0.2$, a warped disk of parent bodies extending only to 50 AU creates an observed warp at 85 AU. In Fig. \ref{fig:planetb}, we plot curves of $a_p$ vs. $m_p$ that produce a warp at 85 AU (eqn. \ref{eqn:constrain}) for $\beta=0$ and $\beta=0.2$. Dust-generating collisions affect the parent bodies' orbits by damping $i_{\rm free}$. However, the largest parent bodies in the collisional cascade, which recently experienced their first collision and have not had $i_{\rm free}$ significantly damped, set the maximum vertical extent of the warp.

\section{Planet b possibly aligned with inner disk}
\label{sec:sculpt}
The framework in Section \ref{sec:model} allows comparison of the model to the observed warp morphology. In creating Fig. \ref{fig:warp}, we used the observed $m_p$ and $a_p$. We selected $i_p = 3.6^\circ$ , corresponding (eqn. \ref{eqn:ip}) to a maximum vertical extent of $z_{\rm warp} = 11$ AU at $a_{\rm warp} = 85$ AU (\citealt{2000H} Fig. 8; \citealt{2006G} Fig. 5). We plot our simulations from the perspective such that both the outer disk and the planet's orbit are perfectly edge-on, as observed, which constrains the orientation of the warp. We find that the distribution of planetesimals matches the disk shape well.

\citet{2011C} reported that planet b is aligned with the flat outer disk and misaligned with the inclined inner disk. Indeed, the difference between two separately measured angles, (a) the intersection of the planetÕs orbit with the sky plane, $31.32^\circ [30.56, 32.12]$ \citep{2011C}, and (b) the sky position of the outer disk, 30-31$^\circ$ \citep{1995K} or $29.5^\circ \pm 0.5$ \citep{2009B}, is consistent with alignment. However, at $\sim2$-$\sigma$, $\beta$ Pictoris b may be misaligned with the flat outer disk by $i_p = 3.6^\circ$, and aligned with the middle of the inclined inner disk, as in our model (Fig. \ref{fig:warp}).

\section{Planet b prevents another planet from creating the warp}
\label{sec:planetc}
If, contrary to the scenario of Section \ref{sec:sculpt}, more precise measurements confirm that planet b's orbit is aligned with the flat outer disk, an undiscovered planet c would be an obvious suspect. However, the presence of planet b severely restricts the parameter space for an additional, warp-making planet. The following loose, generous constraints on planet c must be simultaneously satisfied:

\begin{enumerate}
\item To have escaped radial velocity detection by \citet{2006GL}, planet c must satisfy $$\frac{m_c}{m_{\rm Jupiter}} <  9 \sqrt{a_c/{\rm 1 AU}}$$.
\item Planets b and c must be sufficiently separated for stability, obeying $$\frac{\Delta a}{a} >  2.4 (\mu_b+\mu_c)^{1/3}$$ where $\mu$ is the planet-to-star mass-ratio \citep{1993G}. Even if planets b and c underwent scattering, it is unlikely that their unstable configuration would last long enough to create the warp.
\item Planet c must create the observed disk morphology -- an inclined inner disk from 40-90 AU -- without exciting planet b's inclination. We generously allow any parameters for planet c that produce a forced inclination $i_{\rm forced}<2^\circ$ for planet b and $3^\circ<i_{\rm forced}<8^\circ$ for the inner disk. We calculate the forced inclinations using multi-planet secular theory  \citep[][Section 7.4]{2000M}.
\item Given the system age of 8-20 Myr \citep{2001Z}, planet c must have a mass and semi-major axis small enough so that the evolution time of the warp at 85 AU is slower than 1 Myr (but faster than 20 Myr). If the observed dust grains have $\beta > 0$ (Section \ref{subsec:constrain}), the evolution time must satisfy this constraint at 85 AU $(1-2\beta)$, a more restrictive lower limit. The lower limit is generous, allowing for the possibility that planet c became inclined very recently. 
\item A secular resonance should not disrupt the flat outer disk (90-200 AU); we calculate the locations of secular resonances using multi-planet secular theory.
\end{enumerate}

\begin{figure}[htbp]
\begin{centering}
\includegraphics{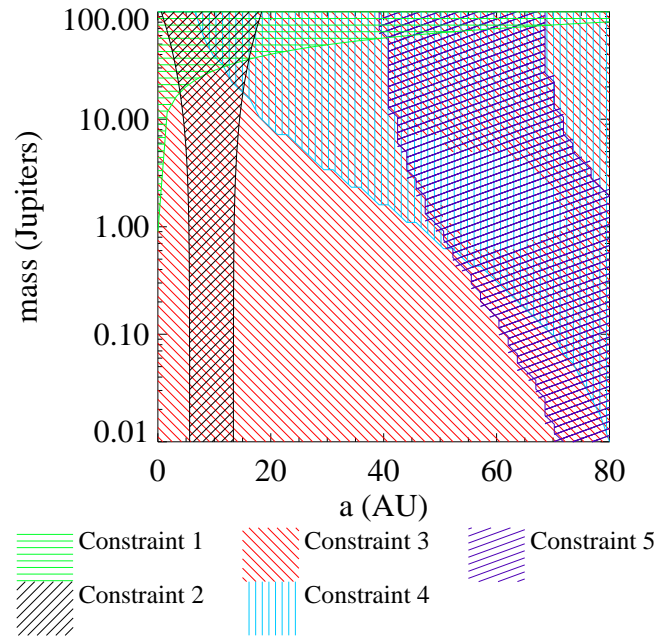}
\caption{Constraints on $a_c$ and $m_c$. The region shaded horizontal-striped green violates Constraint 1 (lack of RV detection), upward-slanted black violates Constraint 2 (stability), downward-slanted red violates Constraint 3 (produces disk morphology without exciting planet b), vertical-striped blue violates Constraint 4 (timescale consistency), and shallow-slant purple violates Constraint 5 (secular resonances in the outer disk).  See text for details. \label{fig:constraints}}
\end{centering}
\end{figure}

We explore the parameter space of planet c's semi-major axis, mass, and inclination. For each combination, we evaluate the equations for the five criteria above in a two-planet system containing planets c, and b, with its nominal $m_b = 9 m_{\rm Jupiter}$, $a_b = 9.5$ AU, and $i_b = 0$. In Fig. \ref{fig:constraints}, we shade the ($a_c$, $m_c$) regions for which no possible inclination of planet c can satisfy the constraints. Other choices for $m_b$ and $a_b$, within observational errors, yield qualitatively similar results. Constraint 3 is most restrictive: the planet must have high enough mass to excite the warp, but low enough mass not to excite $i_b$;  it must be far enough out that the warp can extend to 85 AU, but close enough to incline the inner disk at 40 AU. Constraint 4 considers the time dependence: the warp must reach 85 AU in the system age. The secular oscillation frequency at 85 AU depends on the mass and semi-major axis of both planets b and c -- but not on their inclinations. Planet b, even on a non-inclined orbit, makes a large contribution to this frequency, leaving little room for a contribution from planet c.

Therefore we can rule out that an undiscovered planet c causes the warp, because it cannot do so in the presence of planet b. Additional planets may be present in the system, but are unlikely to be predominantly responsible for the warp.

\section{Planet b's inclination may have damped}
\label{sec:damp}

We demonstrated that planet b's current orbit is consistent with producing the warp only at the extremes of the uncertainties (Section \ref{sec:sculpt}) and that planet b's presence inhibits another planet from creating the warp (Section \ref{sec:planetc}). If follow-up observations confirm the nominal orbit of $\beta$ Pictoris b and position angle of the flat outer disk, we are left uncomfortably with a planet misaligned with the warp it produced. However, we need not regard that conclusion with such discomfort. A sculpted planetesimal disk can record the history of a planet's orbit, revealing a dynamical past we would never guess from the planet's current orbit. For example, Neptune has a nearly circular orbit today but may have have sculpted the Kuiper Belt, our solar system's remnant planetesimal disk, during a period of high eccentricity, which was subsequently damped by dynamical friction \citep[see][and references therein]{2008L}. Embedded in a planetesimal disk, $\beta$ Pictoris b would experience damping of its inclination, though the extent and timescale depend on disk properties.

\subsection{Consistency with disk morphology}

\begin{figure*}[htbp]
\begin{centering}
\includegraphics{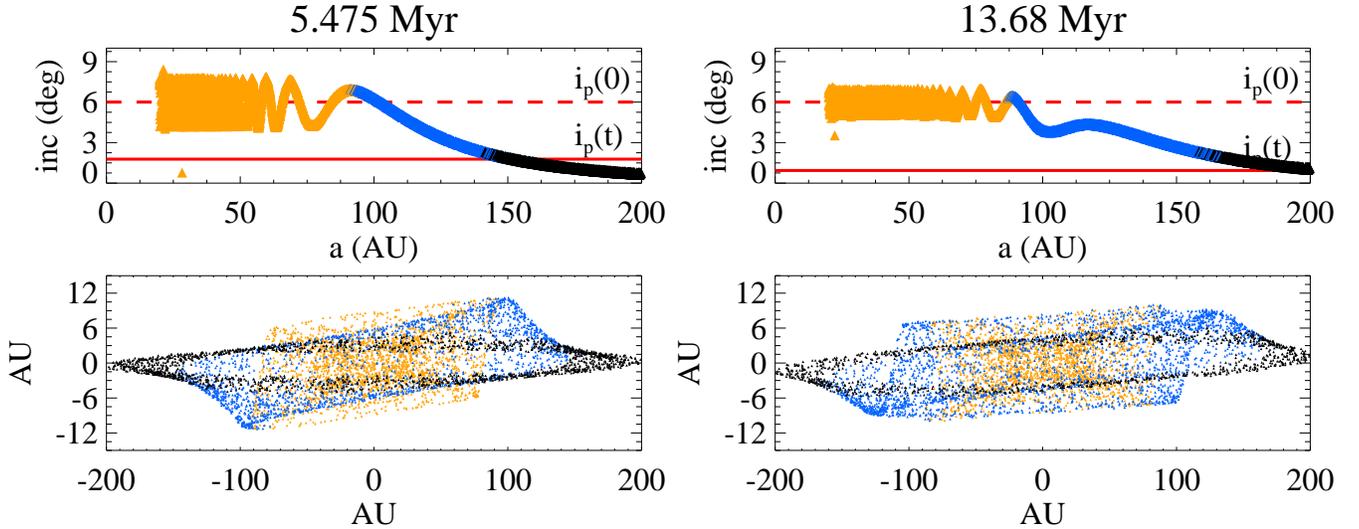}
\caption{Snapshots at two times (left and right) from an N-body simulation of a planet ($m_p = 12 m_{\rm Jupiter}$; $a_p = 13.5$ AU;  initial $i_p=6^\circ$) sculpting an initially flat planetesimal disk. We imposed damping of the planet's inclination of the form $\frac{\dot{i}}{i} = 2\pi$/(4 Myr), following Appendix A of \citet{2011W}. Top: Planetesimal inclinations vs. semi-major axis. The red dashed and solid lines are the planet's initial and current inclination, respectively. Bottom: Projected positions of the planetesimals. See Fig. \ref{fig:warp} for color coding.\label{fig:damp}}
\end{centering}
\end{figure*}

First we demonstrate that the disk morphology can be consistent with the damping of $\beta$ Pictoris b's orbital inclination. The planetesimals, with initial forced inclinations of $i_p$, begin to oscillate about the forced plane and, as we showed in \citet{2011D}, effectively freeze at the inclination values they reach after one damping timescale of the planet's inclination. Therefore, the warp freezes at the distance it reaches when the planet's inclination damps. Because their forced inclinations are damped, the planetesimals have a maximum $i = i_p$ instead of $2 i_p$, requiring $\beta$ Pictoris b to have an initial $i_p \sim 7^\circ$. Fig. \ref{fig:damp}, left panel shows an example: the disk morphology still matches the observations even though the inclined inner disk is not aligned with the planet's orbit. 

However, if the disk evolves for too long after the planet's inclination damps, the planetesimal nodes randomize and the morphology becomes boxy (right panel of Fig. \ref{fig:damp}). The distinction between a consistent and an excessively boxy morphology is qualitative and may require sophisticated modeling of the observed scattered light. However, it is clear that no more than a partial precession period at 85 AU can have passed since the planet's inclination damped. Since the system is young, this requirement does not demand implausible fine-tuning.

\subsection{Damping conditions}

We consider the feasibility of damping planet b's orbital inclination. We expect that planet b started in the outer disk's plane, was perturbed, and is returning to its original plane. Possible perturbations include a second planet scattered inward (e.g. \citealt{2008J}), or resonant-induced inclinations in the disk near the planet (e.g. \citealt{2003T}). 

Ejection of a scattered planet could alter the total angular momentum, requiring a disk with mass comparable to planet b to keep the forced plane, to which planet b damps, low (e.g. an unusually long-lived gaseous proto-planetary disk -- \citealt{1993H} -- or a particularly massive planetesimal belt). We confirmed that fast precession of planet b caused by a massive disk, since the precession occurs about a misaligned axis, would not prevent the excitation of the warp. On longer timescales, the damping medium must allow planet b to remain inclined for almost the system's age.  High-mass planets whose inclinations bring them several scale-heights out of the disk fail to open a gap, and thus gas interaction damps them after only $\sim 100$ orbits \citep{2009M}.  A more moderate inclination ($i\lesssim10^\circ$) for planet b, allowing it to maintain a clean gap, would lead to satisfactorily-inefficient damping \citep{2011B}, perhaps on a 10 Myr timescale.

However, a modest disk mass may suffice. During planet-planet scattering, the total angular momentum is conserved, producing an average forced-plane still aligned with the flat outer disk. Planetesimals contributing to dynamical friction follow the forced plane, with each planetesimal attempting to damp the planet to the planetesimal's own plane. Even if planet b dominates the forced plane in its immediate vicinity, distant planetesimals, within a factor of several of the planet's semi-major axis, could damp the planet's orbit to their flat forced plane (i.e. the plane of the outer disk). (However, planet b would need to dominate the forced plane from 40 - 90 AU to excite the inclined inner disk.) Following \citet{2007F}, Section 2.3, a surface density as low as 3 lunar masses within 40 AU could damp $i_p$ from 10$^\circ$ over 4 million years if the disk remains thin. Collisional dissipation could keep the disk thin \citep[][eqn. 33, 50]{2004G} if
\begin{eqnarray}
\frac{s}{\rm 1cm} < \frac{<i_{\rm disk}>}{2^\circ}\frac{m_{\rm disk}}{5 m_{\rm Earth}}
\end{eqnarray}
where $s$ is a typical planetesimal radius. Since the collisional dissipation rate is $\propto \sigma \propto a^{-1}$, where $\sigma$ is the disk surface density, collisional dissipation could keep the disk thin near planet b while allowing excitation from 40-90 AU, where the warp is created.

Clearly further study is required to find evolutionary scenarios that produce the inclination damping used in Fig. \ref{fig:damp}.  These details are independent of this section's main message: a transient planetary orbit could establish a warp in a disk, to be observed at present, even if that planetary orbit has since changed.

\section{Conclusion}
\label{sec:discuss}

We explored three scenarios for the apparent misalignment between the warped inner disk of $\beta$ Pictoris and the orbit of the directly imaged planet b. In the first, most plausible scenario (Section \ref{sec:sculpt}), planet b's orbit is consistent with producing the warp, at the extremes of the uncertainties. We argued that the alignment depends not only on the planet's orbit but on the (separately measured) position angle of the outer disk. Therefore both of these quantities must be measured more precisely and, if possible, simultaneously from the same images.

In the second, most obvious scenario (Section \ref{sec:planetc}), another planet warps the disk. However, we demonstrated that planet b inhibits another planet from producing the warp. Other planets may exist in the system, creating other disk features, but they cannot be responsible for the warp.

If the first scenario is ruled out by more precise observations, we are left with the third scenario (Section \ref{sec:damp}): planet b created the warp and then had its inclination damped. Detailed modeling of scenarios that allow for the damping of planet b's inclination will be necessary. Confirmation of the damping scenario, especially if observers discover more systems with planets misaligned with the warp they produced, could shed light on disk properties that are important for planet formation but difficult to measure directly.

\acknowledgements We thank Michael Fitzgerald, Paul Kalas, and Philippe Thebault for $\beta$ Pictoris insights, and Thayne Currie and an anonymous referee for helpful comments. R.I.D. acknowledges support by NSF Graduate Research Fellowship DGE-1144152 and D.C.F. by NASA Hubble Fellowship HF-51272.01. Simulations were run on the Odyssey cluster supported by the Harvard FAS Sciences Division Research Computing Group.

 \bibliographystyle{apj}

\end{document}